\UseRawInputEncoding
\documentclass[10pt,conference]{IEEEtran}
\usepackage{cite}
\usepackage{amsmath,amssymb,amsfonts}
\usepackage{algorithmic}
\usepackage{graphicx}
\usepackage{textcomp}
\usepackage[hyphens]{url}
\usepackage{fancyhdr}
\usepackage{hyperref}
\newcommand{\hpcayear}{2024}
\usepackage{ulem}

\usepackage{subfig}
\usepackage[dvipsnames]{xcolor}
\usepackage{pifont}
\usepackage{ulem}

\usepackage{booktabs}
\usepackage{makecell}
\usepackage{array}
\usepackage{multirow}
\usepackage[export]{adjustbox}
\usepackage{tabularx}
\usepackage[ruled,linesnumbered]{algorithm2e}
\usepackage{pifont}

\usepackage{bbding}
\usepackage{listings}
\newcolumntype{Y}{>{\centering\arraybackslash}X}
\newcolumntype{C}[1]{>{\centering\arraybackslash}p{#1}}
\title{An Optimizing Framework on MLIR for Efficient FPGA-based Accelerator Generation}

\def\hpcacameraready{} 

\newcommand\hpcaauthors{Weichuang Zhang\IEEEauthorrefmark{1}, Jieru Zhao\IEEEauthorrefmark{1}\IEEEauthorrefmark{2}, Guan Shen, Quan Chen, Chen Chen, Minyi Guo\IEEEauthorrefmark{2}}
\newcommand\hpcaaffiliation{Department of Computer Science and Engineering, Shanghai Jiao Tong University}
\newcommand\hpcaemail{\{1064080006, zhao-jieru, shenguan\}@sjtu.edu.cn, 
\{chen-quan, chen-chen, guo-my\}@cs.sjtu.edu.cn}

\author{
  \ifdefined\hpcacameraready
    \IEEEauthorblockN{\hpcaauthors{}}
      \IEEEauthorblockA{
        \hpcaaffiliation{} \\
        \hpcaemail{}
      }
  \else
    \IEEEauthorblockN{\normalsize{HPCA \hpcayear{} Submission
      \textbf{\#\hpcasubmissionnumber{}}} \\
      \IEEEauthorblockA{
        Confidential Draft \\
        Do NOT Distribute!!
      }
    }
  \fi 
}

\fancypagestyle{camerareadyfirstpage}{%
  \fancyhead{}
  
  \fancyhead[C]{
    \ifdefined\aeopen
    \parbox[][12mm][t]{13.5cm}{\hpcayear{} IEEE International Symposium on High-Performance Computer Architecture (HPCA)}    
    \else
      \ifdefined\aereviewed
      \parbox[][12mm][t]{13.5cm}{\hpcayear{} IEEE International Symposium on High-Performance Computer Architecture (HPCA)}
      \else
      \ifdefined\aereproduced
      \parbox[][12mm][t]{13.5cm}{\hpcayear{} IEEE International Symposium on High-Performance Computer Architecture (HPCA)}
      \else
      \parbox[][0mm][t]{13.5cm}{\hpcayear{} IEEE International Symposium on High-Performance Computer Architecture (HPCA)}
    \fi 
    \fi 
    \fi 
    \ifdefined\aeopen 
      \includegraphics[width=12mm,height=12mm]{ae-badges/open-research-objects.pdf}
    \fi 
    \ifdefined\aereviewed
      \includegraphics[width=12mm,height=12mm]{ae-badges/research-objects-reviewed.pdf}
    \fi 
    \ifdefined\aereproduced
      \includegraphics[width=12mm,height=12mm]{ae-badges/results-reproduced.pdf}
    \fi
  }
  \fancyfoot[C]{}
}
\begin{document}
\maketitle

\ifdefined\hpcacameraready 
  \thispagestyle{camerareadyfirstpage}
  \pagestyle{empty}
\else
  \thispagestyle{plain}
  \pagestyle{plain}
\fi

\newcommand{\hpcaheight}{0mm}
\ifdefined\eaopen
\renewcommand{\hpcaheight}{12mm}
\fi

\renewcommand{\thefootnote}{}
\footnotetext{\IEEEauthorrefmark{1}The authors contributed equally to this work.}
\footnotetext{\IEEEauthorrefmark{2}Jieru Zhao and Minyi Guo are the corresponding authors.}


\begin{abstract}

With the increasing demand for computing capability given limited resource and power budgets, it is prominent to deploy applications to customized accelerators like FPGAs. However, FPGA programming is non-trivial. 
Although existing high-level synthesis (HLS) tools improve productivity to a certain extent, they are limited in scope and capability to support sufficient FPGA-oriented transformations and optimizations.

This paper focuses on FPGA-based accelerators and proposes POM, an end-to-end optimizing framework built on multi-level intermediate representation (MLIR). POM has several features which demonstrate its scope and capability of performance optimization. First, most HLS tools depend exclusively on a single-level IR like LLVM IR to perform all the optimizations, introducing excessive information into the IR and making debugging an arduous task. In contrast, POM explicitly introduces three layers of IR to perform operations at suitable abstraction levels, streamlining the implementation and debugging process and exhibiting better flexibility, extensibility, and systematicness. Second, POM integrates the polyhedral model into MLIR and hence enables advanced dependence analysis and a wide range of FPGA-oriented loop transformations. By representing nested loops with integer sets and maps at suitable IR, loop transformations can be conducted conveniently through a series of manipulations on polyhedral semantics.
Finally, to further relieve design effort, POM is equipped with a user-friendly programming interface (DSL) that allows a concise description of computation and includes a rich collection of scheduling primitives. An automatic design space exploration (DSE) engine is also provided to search for high-performance optimization schemes efficiently and generate optimized accelerators automatically. 
Experimental results show that POM achieves a $6.46\times$ average speedup on typical benchmark suites and a $6.06\times$ average speedup on real-world applications compared to the state-of-the-art.

\end{abstract}

\section{Introduction}

With the rapid growth of computation-intensive applications,
FPGA-based accelerators are becoming increasingly popular to achieve high processing speeds given limited resource and energy budgets. 
However, programming on FPGAs is challenging. 
To improve productivity, high-level synthesis (HLS) tools are proposed to synthesize behavioral descriptions specified in high-level languages (such as C/C++) into dedicated hardware accelerators \cite{cong2011high}. This allows designers to focus on the behavioral implementation of algorithms without dealing with complex and error-prone digital design. However, there are new challenges to overcome. While existing tools such as Xilinx Vitis HLS offer pragmas to guide the hardware code generation, the quality of generated accelerators largely depends on the user's ability to select appropriate HLS pragmas \cite{zhao2017comba,comba-tcad}. In many cases, it is necessary to restructure source code manually to loosen tight dependencies and achieve high parallelism \cite{zuo2013improving}. This process can be complex and iterative, requiring careful consideration and experimentation, which exacerbates programming and optimization difficulties.

Recent years have witnessed multiple compilation frameworks that cope with programming and optimization difficulties for different platforms, including \ding{172} general processors like CPUs and GPUs, \ding{173} specialized processors with pre-defined hardware like DSAs, and \ding{174} FPGA-based accelerators where data paths can be fully customized and reconfigured. Optimization techniques vary for different hardware and we classify representative frameworks into two categories:

\noindent \textbf{Frameworks for non-FPGA back-ends (\ding{172}\ding{173}):} There is a growing trend towards optimizing applications written in scheduling languages on CPUs, GPUs \cite{ragan2017halide, chen2018tvm, 2015pencil, bondhugula2008pluto, mullapudi2015polymage, baghdadi2019tiramisu, yuki2012alphaz} and specialized processors like DSAs~\cite{moreau2019hardware, 2021akg, 2022exo}. Halide \cite{ragan2017halide} proposes a domain-specific language (DSL) with decoupled computation and schedule for image processing. TVM \cite{chen2018tvm} extends Halide DSL and proposes an optimizing compiler for deep learning. Both of them mainly focus on GPU acceleration. VTA \cite{moreau2019hardware} works as a back-end for TVM and optimizes tensors on an ISA-based DSA processor with pre-defined architecture and FPGA is utilized for prototyping. This is different from commonly used FPGA-based accelerators where the data path is fully customized and reconfigured. Exo \cite{2022exo} provides an abstract programming model for DSAs, which allows developers to write libraries for emerging accelerators. 
Simultaneously, \textit{polyhedral techniques} \cite{2014scheduletree} have shown success in efficient loop analysis and transformation. 
PENCIL \cite{2015pencil}, Pluto \cite{bondhugula2008pluto}, PolyMage~\cite{mullapudi2015polymage}, Tiramisu~\cite{baghdadi2019tiramisu}, and AlphaZ~\cite{yuki2012alphaz} take high-level languages or DSLs as input and generate optimized code automatically for CPUs/GPUs. AKG \cite{2021akg} utilizes polyhedral schedulers and accelerates tensors for NPUs. 
Despite the good performance, none of these works target FPGA-based accelerators, and their loop optimization strategies cannot be adopted directly to FPGAs. Take a nested loop as an example, their strategies seek to parallelize outer loop levels for multi-thread computation and process inner loop levels sequentially within each thread. For AI applications on GPUs and NPUs, loop fusion is utilized to improve data locality and reduce the cost of kernel launch. In contrast, FPGA-friendly optimization strategies tend to pipeline outer loop levels and parallelize inner loop levels through unrolling, and loop fusion mainly reduces resource usage.








\noindent \textbf{Frameworks for FPGA accelerators with customized data paths (\ding{174}):}
Recent advances in HLS frameworks have explored optimization methods for FPGAs \cite{lai2019heterocl,2017halidehls,2020heterohalide,xiang2022heteroflow,ye2022scalehls}. Halide-HLS \cite{2017halidehls} and HeteroHalide \cite{2020heterohalide} work as FPGA back-ends for Halide and generate customized pipelines for image processing. HeteroCL \cite{lai2019heterocl} and HeteroFlow \cite{xiang2022heteroflow} extend TVM DSL and generate spatial architectures on FPGAs. However, these frameworks have limited capabilities in dependence analysis and loop transformation, resulting in a reduced ability to exploit parallelism. All of these works use a single loop-level intermediate representation (IR) for HLS optimization. However, there exist many other schedule methods, and each corresponds to different optimization granularity, which should be applied at or across different abstraction levels for better performance. 
Depending exclusively on a single IR for all optimizations may introduce excessive information into the IR and make debugging an arduous task. 
In contrast, using multiple layers of IRs to represent schedule methods streamlines the implementation process of different analysis, transformation, and optimization methods, which exhibits greater flexibility and systematicness and allows for more efficient design space exploration. 
Most related tools with multi-level IRs target CPUs, GPUs, and specialized processors \cite{baghdadi2019tiramisu, mullapudi2015polymage, gysi2021domain, chelini2021progressive, 2021akg}. ScaleHLS \cite{ye2022scalehls} proposes an HLS framework for FPGAs on top of the multi-level intermediate representation (MLIR) compiler infrastructure~\cite{lattner2020mlir}. It receives C code and expands MLIR with a back-end to generate synthesizable HLS code. However, critical schedule methods and strategies cannot be supported, leading to non-optimal accelerators. Moreover, since the input is C code, designers still need to fully restructure the source code even if the schedule is slightly adjusted.


In this paper, we present \textit{POM}, an open-source optimizing framework on MLIR, which generates efficient FPGA accelerators automatically. POM explicitly divides the compilation process into three layers with hybrid IRs, namely dependence graph IR, polyhedral IR, and annotated affine dialect, which enable various optimizations at appropriate abstraction levels. The dependence graph IR is used for advanced dependence analysis at the graph level of applications. The polyhedral IR is designed to reduce the implementation effort for a wide range of schedule strategies. And the annotated MLIR affine dialect explicitly represents HLS pragmas in loop hierarchies, working as a suitable bridge between polyhedral semantics (the previous layer) and synthesizable HLS code (the output). 
\textcolor{black}{Our main contributions are summarized as follows:}
\vspace{-0.08cm}
\begin{itemize}
\item \textcolor{black}{\textbf{Programmability:} POM provides a decoupled DSL that enables concise descriptions of functions, loops, and arrays. A rich collection of scheduling primitives is provided for flexible customization, leading to much fewer lines of code while maintaining high performance.}
\item \textcolor{black}{\textbf{Extensibility:} POM explicitly introduces three layers of IR to perform operations at suitable abstraction levels in a unified framework, streamlining the implementation and debugging process and reducing the effort of supporting various optimization methods.}
\item \textcolor{black}{\textbf{Quality:} POM provides a rich set of optimization methods and performs FPGA-oriented schedule operations at proper levels, relieving tight loop-carried dependence, exploiting parallelism, and improving overall performance.} 
\item \textcolor{black}{\textbf{Automation:} POM contains a design space exploration (DSE) engine to search for high-performance schedule schemes automatically and efficiently, while also allowing designers to set user-specified schedules.}



\end{itemize}                                                        
Experimental results show that POM achieves $6.46\times$ average speedup on typical benchmark suites and $6.06\times$ average speedup on real-world applications compared to SOTA \cite{ye2022scalehls}.
\section{Background and motivation}\label{sec-background}

\subsection{Polyhedral semantics and dependence analysis}

\begin{figure}
    \centering
    \includegraphics[width=0.48\textwidth]{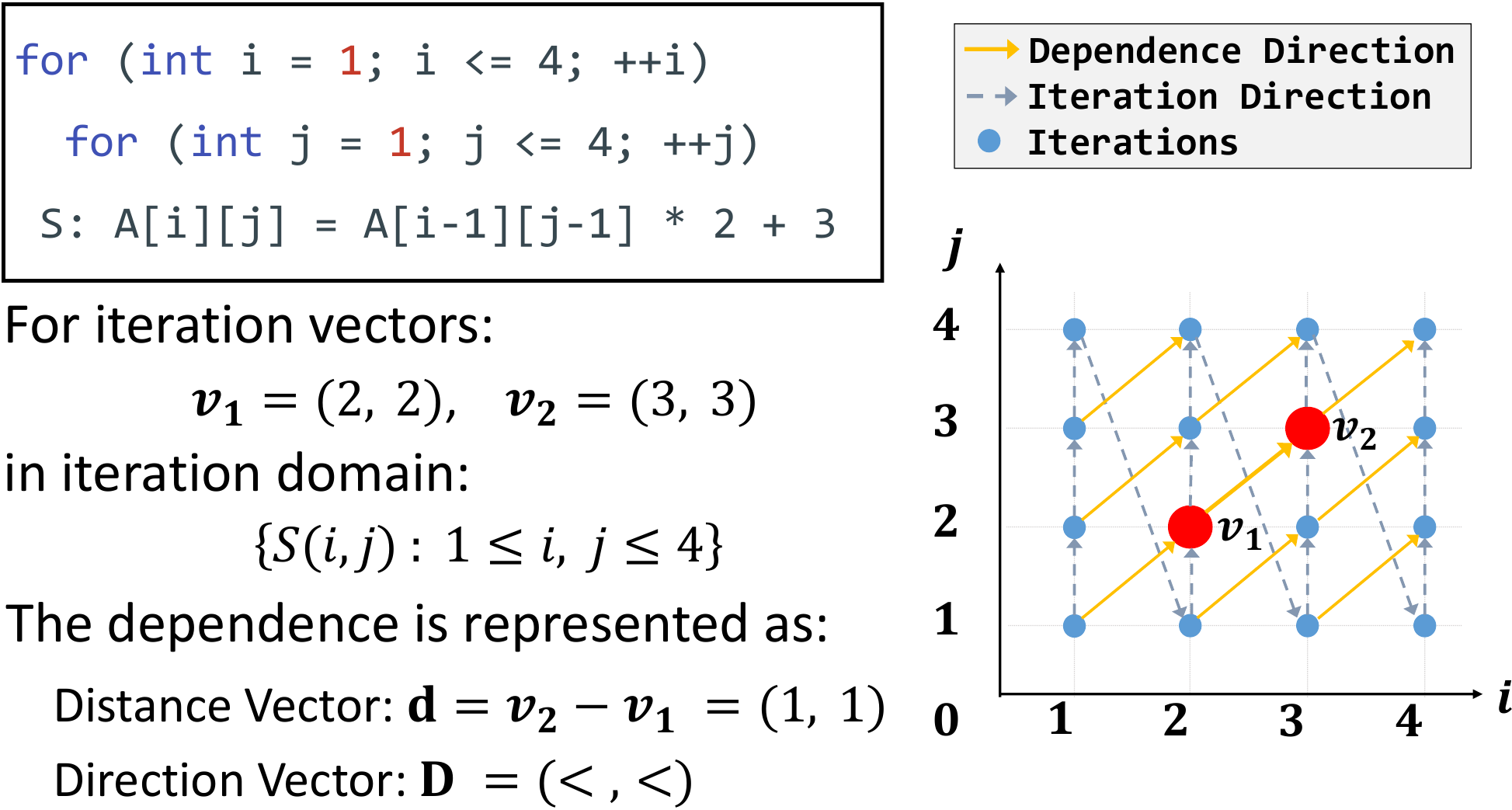}
    \caption{Illustration of loop dependence analysis}
    \vspace{-0.5cm}
    \label{fig:distance-vector}
\end{figure}

The polyhedral model is a powerful compilation technique that represents programs as polyhedrons. These polyhedrons are analyzed and transformed to facilitate program acceleration. Generally, polyhedral semantics extracted from a nested loop can be expressed as an \textit{iteration domain}, \textit{data accesses}, \textit{schedules}, and \textit{dependencies} \cite{kong2013polyhedral}. 
Take the nested loop in Fig. \ref{fig:distance-vector} as an example. \textit{Iteration domain} refers to the set of all statement instances satisfying the loop-bound constraints and is visualized in an n-dimensional iteration space. We can represent a statement instance of an n-level loop as an \textit{iteration vector} with n entries, each of which corresponds to each loop iterator, e.g., $v_1=(2, 2)$ in Fig. \ref{fig:distance-vector}. \textit{Data access} is the affine representation of memory references and \textit{schedule} determines the execution order of statement instances \cite{elango2018diesel}. \textit{Dependence} reflects the data dependence between statement instances. Given two dependent statement instances, their dependence relationship is represented using \textit{distance vector} and \textit{direction vector}. As shown in Fig. \ref{fig:distance-vector}, the \textit{distance vector} \textbf{d} denotes the distance between the source iteration vector $v_1$ and the sink iteration vector $v_2$, and each entry $\text{d}_k$ of \textbf{d} is computed as $(v_2)_k-(v_1)_k$, i.e., $\textbf{d}=(1, 1)$. The \textit{direction vector} has three representations at each entry: $<$, $=$, and $>$, based on whether the corresponding entry of the distance vector is larger than, equal to, or less than zero, correspondingly. In this example, the direction vector \textbf{D} is $(<, <)$. This dependence relationship implies that there exists a loop-carried dependence between loop iterations, which may impede the potential parallelism, especially for FPGA-based accelerators.

\noindent \textbf{Dependence analysis:} POM contains an efficient dependence analysis tool that analyzes distance and direction vectors between dependent statement instances at the \textit{dependence graph IR} level. After identifying implicit data dependence, proper loop transformations can be selected at the following layers to exploit potential parallelism without violating loop dependencies. Details will be introduced in Section \ref{sec-multi-ir}.





\subsection{MLIR Infrastructure}
MLIR \cite{lattner2020mlir} is a compilation stack built on LLVM \cite{lattner2004llvm}. Unlike Clang \cite{clang} and other mature compilers with fixed abstraction levels, MLIR provides multi-level IRs and enables flexible optimization and transformation methods at different IR levels.
MLIR offers a well-defined infrastructure for users to organize values, operations, types, and attributes in  \textit{dialects}. There are dozens of \textit{dialects} in MLIR's ecological system, such as the tensor dialect \cite{tensordialect} for tensor creation and manipulation and the CIRCT dialect \cite{circtdialect} for efficient hardware designs. POM uses a mixture of dialects, including the affine dialect \cite{affinedialect}, the arith dialect \cite{arithdialect}, and the memref dialect \cite{memrefdialect}, to describe the lower-level IR that is converted from our polyhedral IR. 
The loop body and operations like load and store are represented with the affine dialect which provides an abstraction for affine operations and is a suitable IR to which our polyhedral IR is lowered. The arith dialect is intended to perform fundamental arithmetic operations such as binary and ternary arithmetic operations on integer and floating point numbers.
The memref dialect provides a memory reference of arrays and tensors. Details will be introduced in Section \ref{sec-multi-ir}.

\begin{figure*}
    \centering
    \includegraphics[width=\textwidth]{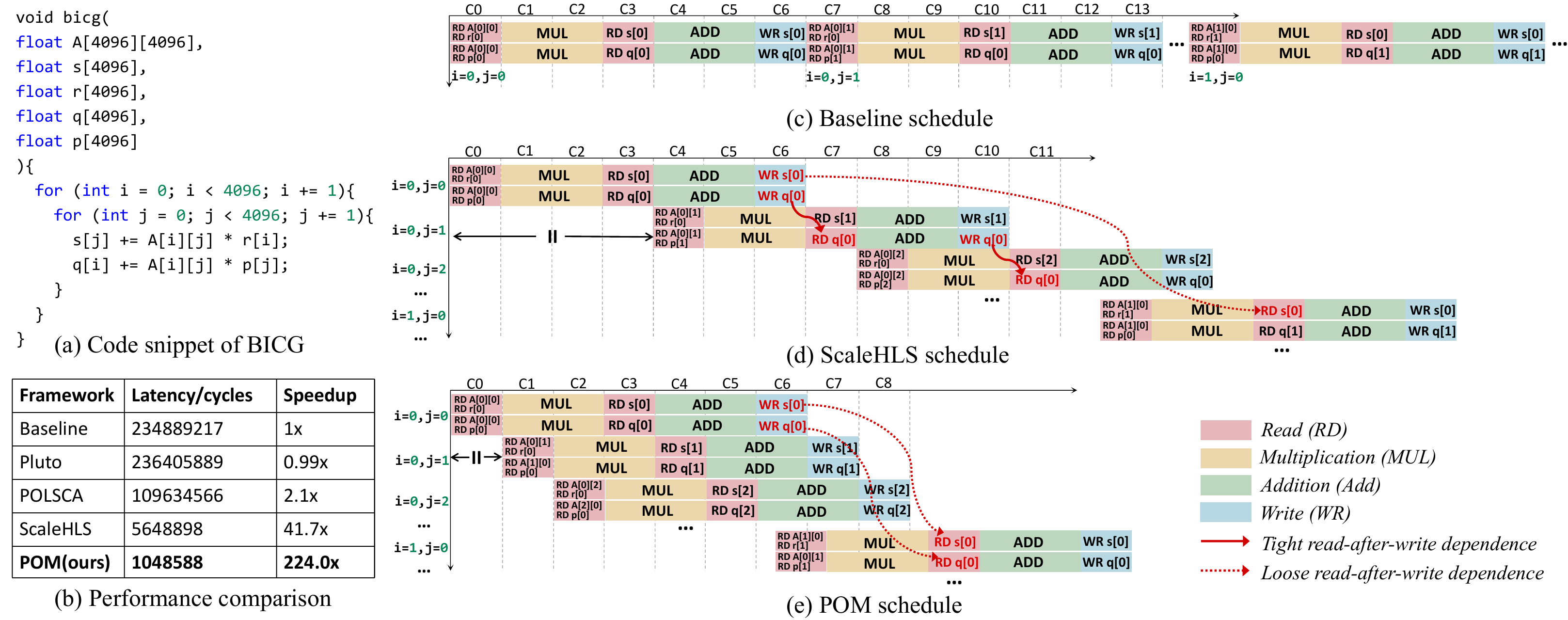}
    \caption{
    \textcolor{black}{Motivating example: (a) presents the code snippet of BICG; (b) compares \textit{latency} and \textit{speedup} achieved by different frameworks; (c)(d)(e) illustrate schedules for BICG generated by the baseline, ScaleHLS, and POM, correspondingly.}}
    \label{fig:motivating}
\end{figure*}

\vspace{-0.1cm}
\subsection{Issues in prior works}
\vspace{-0.1cm}
 
We compare representative and latest frameworks to illustrate existing issues in Table \ref{comparison-table}.
Existing automatic tools like Pluto \cite{bondhugula2008pluto} have shown the power of polyhedral techniques \cite{baghdadi2019tiramisu, bondhugula2008pluto, mullapudi2015polymage,2015pencil,yuki2012alphaz}.
However, none of them target FPGAs and deliver low performance if their schedule strategies are directly reused for FPGA accelerators. 
FPGA-related studies using polyhedral techniques~\cite{2018polysa,2021autosa,zuo2013improving,2022fplpolsca} have limitations in either performance or generality. PolySA \cite{2018polysa} and AutoSA \cite{2021autosa} generate systolic arrays for dense matrices. However, the performance degrades on workloads of which the dependence distance is larger than one, such as the stencil computation \textit{Seidel} \cite{polybenchcode}.
Zuo et al. use the polyhedral model to handle data-dependent modules with stencils in a small design space, limiting its upper bound of optimization \cite{zuo2013improving}. 
POLSCA \cite{2022fplpolsca} introduces Pluto into MLIR and directs Pluto to generate codes that can be consumed by HLS tools. However, it demonstrates limited performance by adopting the Pluto schedule which is better suited for multi-core CPUs. 

\begin{table}[]\large
\renewcommand\arraystretch{1.2}
\caption{Comparison between representative frameworks.}
\label{comparison-table}
\resizebox{\columnwidth}{!}{%
\begin{tabular}{lcccccc}
\hline

\textbf{Feature}     & \textbf{Pluto}   & \textbf{POLSCA} & \textbf{HeteroCL} & \textbf{\begin{tabular}[c]{@{}c@{}}Hetero-\\ Halide\end{tabular}}    & \textbf{\begin{tabular}[c]{@{}c@{}}Scale-\\ HLS\end{tabular}} &  \textbf{POM}         \\ \hline
\textbf{Productivity} \\
Scheduling language & \color{Red}{\ding{56}}  &\color{Red}{\ding{56}}  &  \color{ForestGreen}{\ding{52}}  & \color{ForestGreen}{\ding{52}}  & \color{Red}{\ding{56}} & \color{ForestGreen}{\ding{52}}     \\
Multi-level IR        & \color{Red}{\ding{56}}   & \color{ForestGreen}{\ding{52}}   & \color{Red}{\ding{56}} & \color{Red}{\ding{56}}   & \color{ForestGreen}{\ding{52}} & \color{ForestGreen}{\ding{52}} \\ 
User-specified scheduling         & \color{Red}{\ding{56}}    &  \color{Red}{\ding{56}} &  \color{ForestGreen}{\ding{52}} & \color{ForestGreen}{\ding{52}}  & \color{Red}{\ding{56}}  & \color{ForestGreen}{\ding{52}}   \\
Automated DSE/scheduling     & \color{ForestGreen}{\ding{52}}     & \color{ForestGreen}{\ding{52}}  &  \color{Red}{\ding{56}}   & \color{ForestGreen}{\ding{52}}    & \color{ForestGreen}{\ding{52}}   & \color{ForestGreen}{\ding{52}}  \\
\hline
\textbf{Capability and Efficiency} \\
Polyhedral model    & \color{ForestGreen}{\ding{52}}  & \color{ForestGreen}{\ding{52}}  & \color{Red}{\ding{56}}  & \color{Red}{\ding{56}}   & \color{Red}{\ding{56}} & \color{ForestGreen}{\ding{52}}         \\ 
FPGA-oriented loop transformation   & \color{Red}{\ding{56}}   & \color{Red}{\ding{56}}  & \color{Red}{Limited}   & \color{Red}{Limited}  & \color{Red}{Limited}  & \color{ForestGreen}{\ding{52}}          \\ 
HLS hardware optimization      & \color{Red}{\ding{56}}  & \color{ForestGreen}{\ding{52}}  & \color{ForestGreen}{\ding{52}}  & \color{ForestGreen}{\ding{52}}  & \color{ForestGreen}{\ding{52}}  & \color{ForestGreen}{\ding{52}}       \\ 
Ability of data type customization     & \color{Red}{\ding{56}}    & \color{Red}{\ding{56}} &  \color{ForestGreen}{\ding{52}} & \color{ForestGreen}{\ding{52}}    & \color{Red}{\ding{56}}     &\color{ForestGreen}{\ding{52}}         \\ 
\hline
\textbf{Generality} \\
Apply to multiple domains    & \color{ForestGreen}{\ding{52}}   & \color{ForestGreen}{\ding{52}}  & \color{ForestGreen}{\ding{52}} & \color{Red}{\ding{56}}   & \color{ForestGreen}{\ding{52}}   &  \color{ForestGreen}{\ding{52}}       \\ \hline
\end{tabular}%
}
\vspace{-0.5cm}
\end{table}

Recent advances in scheduling languages for FPGA frameworks also demonstrate the potential to improve productivity. HeteroCL \cite{lai2019heterocl} and HeteroFlow \cite{xiang2022heteroflow} extend TVM DSL and optimize code using Halide IR. Halide-HLS \cite{2017halidehls} and HeteroHalide \cite{2020heterohalide} work as FPGA back-ends of Halide for efficient image processing. 
We compare HeteroCL and HeteroHalide in Table \ref{comparison-table}. They perform several hardware optimizations on a loop-level IR extended from Halide IR. However, they have limited capabilities of dependence analysis and loop transformations based on a single IR.
As discussed before, different schedule methods correspond to different optimization granularity and should be applied at appropriate abstraction levels. Adopting multi-level IRs provides a flexible and systemic way to simplify this implementation process and makes it easier to achieve better performance.

ScaleHLS \cite{ye2022scalehls} proposes the first HLS framework for FPGAs on MLIR to optimize the input HLS C code. However, all the loop-level transformations are performed at MLIR dialects, and it is not convenient to add a new schedule method since the implementation requires modifying IR operations and passes. Therefore, some critical FPGA-oriented schedule strategies are not considered, leading to non-optimal performance. Also, designers cannot freely determine user-specified schedules, limiting the flexibility.
In contrast, POM introduces a new polyhedral IR into MLIR and hence enables advanced dependence analysis and a wide range of FPGA-oriented loop transformations. By representing nested loops as integer sets and maps at our polyhedral IR, loop transformations can be conducted conveniently through a series of manipulations on polyhedral semantics. Moreover, POM lowers IRs progressively by performing proper operations at suitable IR levels, provides a DSL for users to specify a customized schedule, and is applicable to multiple domains, such as image processing, linear algebra, stencils, and deep learning. A DSE engine is also provided to search for a proper schedule automatically.

\vspace{-0.1cm}
\subsection{The motivating example}
\vspace{-0.1cm}

\textcolor{black}{We take \textit{BICG} \cite{polybenchcode} in Fig. \ref{fig:motivating}(a) as a motivating example to illustrate the effects of POM.} We evaluate the performance of the original code without optimization (baseline) and the optimized code generated by four frameworks, namely Pluto \cite{bondhugula2008pluto}, POLSCA \cite{2022fplpolsca}, ScaleHLS \cite{ye2022scalehls}, and POM. The target device is Xilinx XC7Z020 FPGA. The performance varies in \textit{latency} and \textit{speedup}, as shown in Fig. \ref{fig:motivating}(b). We also present corresponding generated schedules in Fig. \ref{fig:motivating}(c)(d)(e). The horizontal axis denotes clock cycles and the vertical axis denotes the execution order of loop iterations. For each iteration, there are two rows to represent the two statements in the loop. Unrolled iterations are omitted for simplification. 

By default, loop iterations are executed sequentially as shown in Fig. \ref{fig:motivating}(c). Pluto enables automatic loop transformation based on the polyhedral model and target CPU/GPU acceleration. Therefore, it focuses on dividing loops into tiles and improving data locality, while parallelizing iterations at outermost loops. This strategy is not suitable for FPGA accelerators where high performance is usually achieved by a deeply pipelined datapath and innermost loops are unrolled for greater parallelism. The generated schedule of Pluto is similar to Fig. \ref{fig:motivating}(c) with slight differences in the execution order of iterations. 
POLSCA utilizes Pluto to perform automatic loop transformations and then conduct several HLS hardware optimizations. However, there still exists loop-carried dependence in the generated code, restricting the parallelism degree. Additionally, it fails to perform proper array partitioning for large sizes like 4096. As a consequence, it results in a schedule with an unsatisfying initiation interval, i.e., $II=167$. 

ScaleHLS attempts to relieve tight loop-carried dependence by performing \textit{loop interchange}. For example, q[i] is written in each iteration and read by subsequent iterations along the j-dimension, incurring a tight loop-carried dependence (highlighted with red solid lines). To solve this issue, \textit{loop interchange} is applied to move the j-level loop to the outermost, enlarging the distance between dependent read and write operations. However, the $II$ between loop iterations cannot be reduced, because s[j] would be written and read by subsequent iterations along the current inner dimension, i.e., i-dimension, after interchanging loop levels. Therefore, ScaleHLS cannot relieve the tight dependence for all statements simultaneously and achieves non-optimal performance. The actual $II$ is 43 considering unrolled iterations. 
POM captures this sophisticated dependence and performs \textit{loop split-interchange-merge} to relieve tight dependence, generating an optimized schedule with $II=2$. As shown in Fig. \ref{fig:motivating}(e), the distances between dependent reads and writes are enlarged (highlighted with red dash lines) and the generated accelerator executes as a highly efficient pipeline. Transformation details are shown in Fig. \ref{fig:bicg}. 

\section{Framework Overview}
Figure 3 depicts a high-level overview of POM. The POM DSL describes the algorithm specification and schedule as input. Then based on our multi-level IR infrastructure, POM first captures data dependencies and generates an optimized data-dependence graph, which is represented as \textit{dependence graph IR}. Next, it extracts polyhedral semantics, perform transformations on polyhedral representations, and yields the \textit{polyhedral IR}. The polyhedral IR is then lowered to MLIR affine dialect with HLS attributes, where HLS hardware optimizations are performed. Finally, the optimized and annotated affine dialect is translated into synthesizable HLS code. Various operations of analysis, transformation, and optimization are performed at different stages. To help users find a high-performance design choice in an extremely large design space, an automatic design space exploration (DSE) engine is integrated into POM. Details will be introduced in corresponding sections.

\section{The programming model}


POM is equipped with a declarative DSL embedded in C++ to describe loop nests, functions, and arrays. Our DSL inherits the idea of Halide \cite{ragan2017halide} and decouples the algorithm specification from the schedule. This decoupled programming model allows users to write an architecture-independent algorithm and specify a set of scheduling primitives that determines the execution order of operations. By setting user-specified scheduling primitives, programmers can explore different transformation and optimization strategies freely without restructuring the code heavily. \textit{Different from Halide DSL, our DSL is well-designed to extract polyhedral semantics easily, while maintaining simplicity and efficiency.} With clear abstractions to represent variables, multi-dimensional arrays, loop nests, and functions, POM DSL is capable of describing a wide variety of computation-intensive algorithms, such as linear algebra, stencils, image processing, and deep learning.

\begin{figure}
    \centering
    \includegraphics[width=0.47\textwidth]{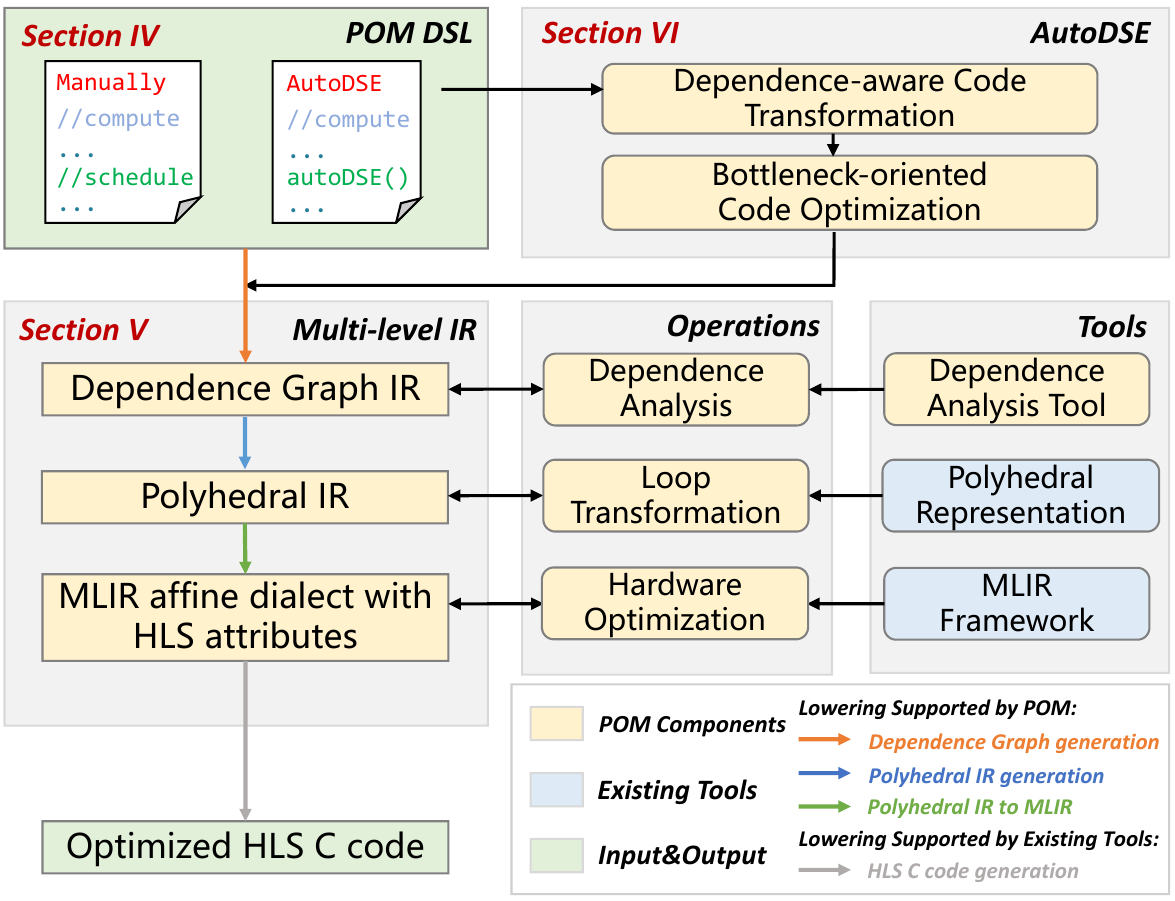}
    \caption{Framework overview}
    \label{fig:my_label}
    \vspace{-0.5cm}
\end{figure}

\subsection{Algorithm specification}

\color{black}POM DSL describes an algorithm specification using the \texttt{compute} operation. Figure \ref{fig:PM_DSL} presents a matrix multiplication kernel described with POM DSL. We first declare names and ranges of loop iterators (L2) and declare three placeholders that represent arrays A, B, and C (L4-L6). Initialization steps are omitted for simplicity. Then we instantiate a \texttt{compute} operation to describe the matrix multiplication algorithm (L8). 
Instead of explicitly writing a loop, programmers can define the iteration domain, the statement, and the destination placeholder for results in a single line. This simplifies the transformation from \texttt{compute} to its polyhedral representation. On the one hand, polyhedral semantics such as iteration domain can be directly obtained from \texttt{compute}; On the other hand, the data dependence is explicitly shown by load and store operations of each \texttt{compute}.
Finally, \texttt{codegen} can be added (L9) to generate the corresponding HLS code.
\color{Black}

Additionally, algorithms implemented with different data types vary in performance on FPGAs. To enable flexible customization, POM supports multiple data types to specify variables and arrays, including signed and unsigned integers with 8, 16, 32, 64 bits, 32-bit single-precision floating-point, and 64-bit double-precision floating-point. Note that our DSL can be easily extended to support more customized data types.

    
    
    

\subsection{Scheduling primitives}
POM automates and simplifies performance optimization 
with a few lines of code specifying scheduling primitives. A rich set of scheduling primitives is provided, as shown in Table \ref{table:looptrans}. Programmers can explore different schedule strategies by instantiating desired scheduling primitives without modifying the algorithm specification. We also provide a primitive \texttt{f.auto\_DSE()} for automatic design space exploration.

\begin{figure}
    \centering
    \includegraphics[width=0.48\textwidth]{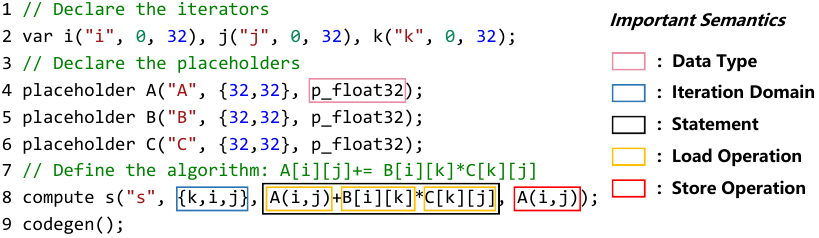}
    \caption{\textcolor{black}{Matrix multiplication with POM DSL.} 
    }
    \label{fig:PM_DSL}
    \vspace{-0.3cm}
\end{figure}

\begin{figure}
    \centering
    \includegraphics[width=0.35\textwidth]{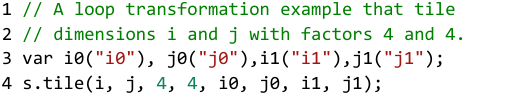}
    \caption{Loop tiling on the algorithm in Fig. \ref{fig:PM_DSL}. }
    \label{fig:PM_tile}
    \vspace{-0.5cm}
\end{figure}

\noindent \textbf{Primitives for loop transformations:} 
Effective loop transformations are necessary to restructure the code and make it better fit with following hardware optimizations. 
Table \ref{table:looptrans} lists transformation primitives we have supported and presents how they work on the iteration domain. For example, \textit{loop skewing} changes the dependence direction by \textit{skewing} the iteration domain.
We continue to use the example in Fig. \ref{fig:PM_DSL} for demonstration. As shown in Fig. \ref{fig:PM_tile}, \textit{loop tiling} is applied using the \texttt{tile} primitive (L4). The dimension \texttt{i} and \texttt{j} are divided into four new dimensions \texttt{i0}, \texttt{j0}, \texttt{i1}, \texttt{j1} with given factors \texttt{(4, 4)}. Note that the rationale of loop tiling for FPGA accelerators is not the same as that for CPUs and GPUs which mainly exploit data localities. Loop tiling here sets proper factors to partition loop levels without loop-carried dependence and move them into inner loop levels, which will be unrolled at the lower IR level. 
\textcolor{black}{
Moreover, determining the execution order of computations is crucial when there exist multiple computations. This requires discerning the dimension of the first computation after which the second one is executed. POM addresses it by organizing the sequence of loops using the \texttt{after} method, providing a fine-grained control.}
\color{black} 

\noindent \textbf{Primitives for HLS hardware optimizations:}
HLS hardware optimizations, represented as HLS pragmas, are supported by POM with a set of primitives.
as shown in Table \ref{table:looptrans}. These primitives will be translated into corresponding HLS pragmas during code generation. We continue to perform hardware optimizations following Fig. \ref{fig:PM_tile}, as shown in Fig. \ref{fig:PM_HLS}. To exploit parallelism, we apply \textit{loop pipelining} at loop level $j0$ (L2) and \textit{unroll} its inner loops $i1$ and $j1$ (L3-L4) completely (setting unroll factor to 4). To guarantee parallel memory accesses, \textit{array partitioning} with suitable partition options and factors is applied to improve the memory performance (L5). The equivalent HLS C code is shown in L7-L18.

\begin{figure}
    \centering
    \vspace{-0.25cm}
    \includegraphics[width=0.43\textwidth]{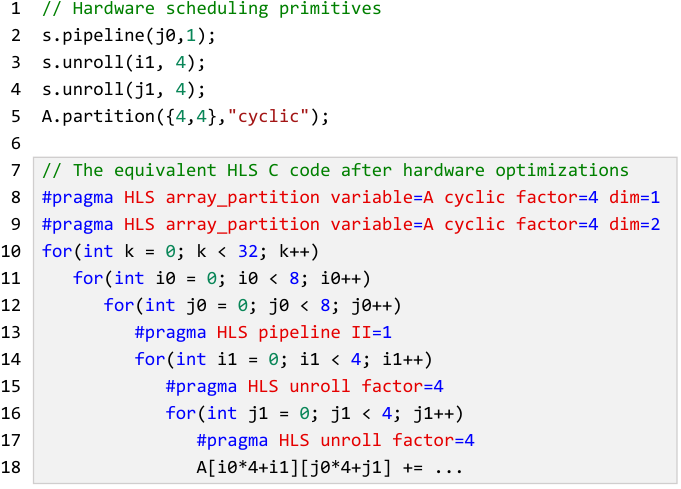}
    \caption{Hardware optimizations and the equivalent HLS code. }
    \label{fig:PM_HLS}
\end{figure}

\begin{figure*}
    \centering
    \includegraphics[width=0.97\textwidth]{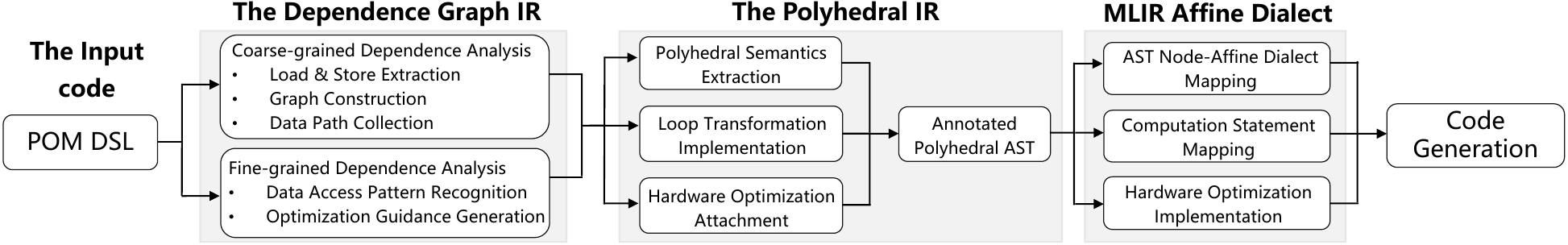}
    \caption{The compilation flow in POM.}
    \label{fig:IR transformation}
    \vspace{-0.4cm}
\end{figure*}

\noindent \textbf{Primitive for automatic design space exploration:}
Despite a series of primitives provided, it requires the programmer's expertise to select a proper combination. Moreover, exploring the huge design space of combinations manually is time-consuming and can easily fall into sub-optimal designs. \textcolor{black}{Considering these issues, POM provides a \texttt{f.auto\_DSE()} primitive, with which programmers can rely on POM to generate high-quality accelerators automatically. Details of DSE will be introduced in Section \ref{sec-auto-dse}. }

\begin{table}[]
\footnotesize
\renewcommand\arraystretch{1.1}
\caption{Scheduling primitives provided by POM.}
  \label{table:looptrans}
  \vspace{-0.1cm}
\begin{tabular}{ll}
\toprule[1pt]
\textbf{Primitive}                                                             & \textbf{Description}                                                                                                                                     \\ \midrule[1pt]
\multicolumn{1}{c}{\textbf{}}                                                  & \textbf{Loop Transformation}                                                                                                                             \\ 
\hline
s.interchange(i, j)                                                            & Interchange loop level i and j of compute s.                                                                                                             \\ 
\hline
s.split(i, t, i0, i1)                                                          & \begin{tabular}[c]{@{}l@{}}Split loop level i of compute s with factor t. \\ The generated loop levels are (i0, i1).\end{tabular}                        \\ \hline
\begin{tabular}[c]{@{}l@{}}s.tile(i, j, t1, t2,\\ \qquad  i0, j0, i1, j1)\end{tabular} & \begin{tabular}[c]{@{}l@{}}Tile loop levels (i, j) of compute s with factors \\ (t1, t2), generating loop levels (i0, j0, i1, j1).\end{tabular} \\ 
\hline
\begin{tabular}[c]{@{}l@{}}s.skew(i, j, t1, t2,\\ \qquad \quad i', j')\end{tabular}     & \begin{tabular}[c]{@{}l@{}}Skew loop levels (i, j) of compute s with \\ factors (t1, t2), generating loop levels (i', j').\end{tabular}           \\ 
\hline
s1.after(s2, j)                                                                 & \begin{tabular}[c]{@{}l@{}}Compute s1 is executed after compute s2 at \\ loop level j.\end{tabular}                                                      \\ 
\hline
\multicolumn{1}{c}{\textbf{}}                                                  & \textbf{Hardware Optimization}                                   \\ 
\hline
s.pipeline(i, t)                                                               & 
\begin{tabular}[c]{@{}l@{}}Pipeline the loop at level i with II = t.\end{tabular}                                                          \\ 
\hline
\begin{tabular}[c]{@{}l@{}}A.partition(\{t1, t2\},\\ \qquad \qquad ``cyclic”)\end{tabular}     & \begin{tabular}[c]{@{}l@{}}Partition the array A with factor t1 at the first\\ dimension and t2 at the second dimension.\end{tabular}                    
\\ 
\hline
s.unroll(i, t)                                                                 & Unroll the loop level i with factor t.                                                                                                                   \\ 
\hline
\multicolumn{1}{c}{\textbf{}}                                                  & \textbf{Design Space Exploration}                                                                                                                        \\ 
\hline
f.auto\_DSE(``PATH")                                                                  & \begin{tabular}[c]{@{}l@{}}Perform design space exploration for function f\\ automatically.\end{tabular}                                                 \\ \bottomrule[1pt]
\end{tabular}
\vspace{-0.3cm}
\end{table}

\section{Multi-level IR in POM} \label{sec-multi-ir}

Figure \ref{fig:IR transformation} shows the complete compilation flow. 
POM explicitly divides the compilation process into three layers with hybrid IRs, lowering the DSL progressively and transforming the code at different abstraction levels with a given schedule. Dependence analysis, code transformation, and hardware optimization are performed at appropriate IR levels, namely dependence graph IR, polyhedral IR, and MLIR affine dialect with HLS pragma attributes.

\begin{figure}
    \centering
    \includegraphics[width=0.49\textwidth]{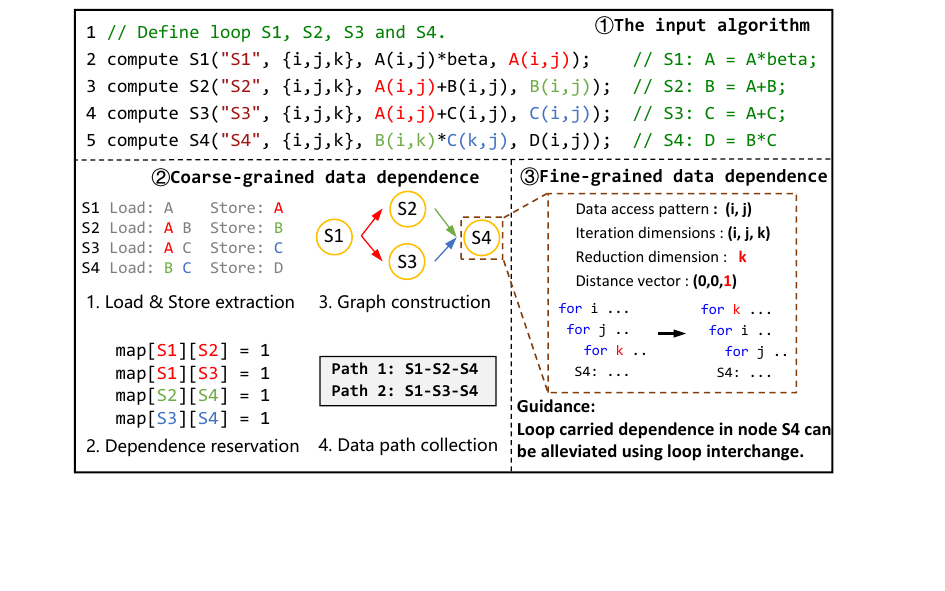}
    \caption{Illustration of dependence graph IR.}
    \label{fig:dependence graph}
    \vspace{-0.5cm}
\end{figure}


\begin{figure*}
    \centering
    \includegraphics[width=\textwidth]{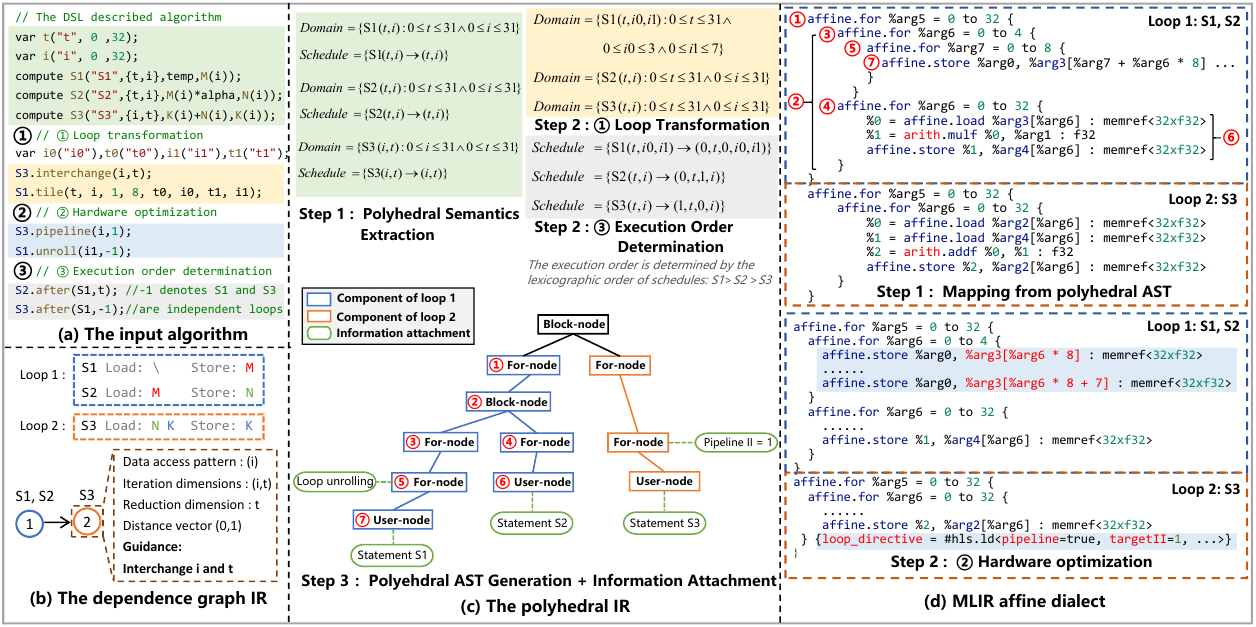}
    \caption{The lowering process of multi-level IR in POM}
    \label{fig:polyhedralir}
\end{figure*}

\vspace{-0.05cm}
\subsection{Dependence Graph IR}
\vspace{-0.1cm}

The initial level of IR in POM is referred to as the dependence graph IR, which facilitates both coarse-grained and fine-grained dependence analyses. 

\noindent \textbf{Coarse-grained dependence analysis}: Figure \ref{fig:dependence graph}\ding{173} illustrates the data dependence graph construction process and the dependence analysis conducted on it. The POM DSL represents explicit producer-consumer relation in the definition of \texttt{compute}, reflecting coarse-grained data dependence between different loops. POM captures these coarse-grained data dependencies between \texttt{computes} by extracting and analyzing the \texttt{load} and \texttt{store} operations and preserves them using a dependence map. With the information from the dependence map, a dependence graph is constructed, where each \texttt{node} represents a nested loop, and each \texttt{edge} signifies the dependence between two loops. POM employs a Depth-First Search (DFS)-based approach to traverse the graph and collect all the \texttt{data paths}, which is critical information for subsequent design space exploration.

\noindent \textbf{Fine-grained dependence analysis}: Once the dependence graph is constructed, fine-grained data dependence analysis is carried out to analyze dependencies between consecutive iterations of loops. These dependencies, known as loop-carried dependencies, can pose challenges to achieving maximum parallelism, particularly in the context of FPGA-based accelerators.
Therefore, it is essential to identify loop-carried dependencies in the dependence graph IR and guide lower-level transformations and optimizations. Specifically, POM traverses each node in the graph, analyzes the dependence by calculating the \textit{distance and direction vectors} of loops, and stores related information as \textit{node attributes}. Figure \ref{fig:dependence graph}\ding{174} illustrates an example of fine-grained data dependence analysis for node S4. POM first captures the data access pattern of the placeholder (D) that stores the updated value, which is (i, j). 
Considering both the data access pattern and iteration dimensions (i, j, k), POM determines the reduction dimension, which in this case is \textit{k}. 
The distance vector for node S4, with iteration dimensions (i, j, k), is then computed as (0, 0, 1), indicating the presence of loop-carried dependence in the \textit{k} dimension. This information is valuable for guiding loop transformations, specifically loop interchange in this example, which involves swapping the inner loop \textit{k} with tight dependencies with the outer loop. Moreover, the identification of loop-carried dependence can serve as a hint to users, directing them to set the HLS DEPENDENCE pragma.

\subsection{Polyhedral IR}\label{sec:polyhedralir}
Once the dependence analysis is complete, the dependence graph IR is lowered to the polyhedral IR, where various loop transformations are implemented at this level. In this section, we discuss the rationale for the polyhedral IR, its construction process, and the loop transformations implemented on it.

\noindent \textbf{Rationale for the polyhedral IR}:
Although the affine dialect does support partial polyhedral loop transformations, we have chosen to introduce an additional polyhedral IR to implement FPGA-friendly loop transformations with the integer sets and maps from Integer Set Library (isl) \cite{verdoolaege2010isl}. We have two considerations.
1) \textit{Efficiency}: as is discussed in the MLIR official document \cite{mlirpoly}, performing specific transformations on integer sets and maps, rather than the entire nested loop structure, offers more simplicity, particularly when dealing with complex cases such as skewing. 2) \textit{Scope}: 
isl library is able to perform various set operations to handle constraints between iteration domains, which makes it capable of generating code for any arbitrary affine schedule, whereas MLIR affine dialect lacks this functionality \cite{mlirpoly2}. This means that the transformations available in the affine dialect have certain restrictions compared to more generic polyhedral models like isl, which can handle any affine schedule. For instance, loop fusion in affine dialect currently only fuse loop nests with single-writer/single-reader dependence with the same constant loop bounds. 

\noindent \textbf{Construction of the polyhedral IR}:
Figure \ref{fig:polyhedralir}(c) illustrates the construction process of our polyhedral IR. In the first step, the polyhedral semantics of each node (nested loop) in the dependence graph IR, such as iteration domain and schedules, are extracted and represented by integer sets and maps efficiently. In the second step, given user-specified primitives (\ding{172}) in DSL, iteration domains are transformed by our pre-implemented loop transformation methods correspondingly. Similarly, schedules are modified to guarantee the execution order of \texttt{computes}, as specified in DSL (\ding{174}). This is based on the lexicographic order theory \cite{2014scheduletree}. In the third step, a union map is created by collecting all the domains and schedules of different loops in one integer map. Then an \textit{ast\_build} method from isl is invoked to build the \textit{polyhedral AST} from the union map.
The generated AST contains four types of nodes: \textit{if-node}, \textit{for-node}, \textit{block-node}, and \textit{user-node}, which can be seamlessly translated to MLIR affine dialect. Additionally, since the \textit{polyhedral AST} lacks representation for computation, we attach critical information such as computation statements to \textit{user-nodes} (\ding{178}). During IR lowering, this information is retrieved to generate computation statements in affine dialect, as shown in Fig. \ref{fig:polyhedralir}(d) (\ding{178}). 
Furthermore, the hardware optimization information is attached to the corresponding node within the AST. This enables the optimization information to be preserved and utilized in the next layer of IR.

\noindent \textbf{Implementation of loop transformations}: By representing nested loops with integer sets and maps at the polyhedral IR level, 
loop transformations can be formulated as a series of manipulations on polyhedral semantics. 
These manipulations include the interchange between loop dimensions, the calculation of new iteration domains through mathematical methods, the modification of array indexes, etc. 
We take the loop tiling process in Fig. \ref{fig:polyhedralir} as an example. Tiling the loop level $i$ with factor 8 will change the iteration domain \{$S(t,i)$\}. The computation process of the new domain is \{$S(t,i0,i1): 0 \leq t \leq 31 \land  i0 = floor(i/8) \land i1 = i \% 8 \land 0 \leq i \leq 31 $\}, namely \{$S(t,i0,i1): 0 \leq t \leq 31 \land 0 \leq i0 \leq 3\land 0 \leq i1 \leq 7$\}.
Note that besides operations on rectangular iteration domains, POM is capable of handling non-rectangular iteration domains through more complex loop transformations such as \textit{loop skewing} with the guidance of dependence analysis.

We implement the most commonly used loop transformations as a library in Table \ref{table:looptrans}. Users only need to specify APIs provided in our DSL and invoke POM to perform automatic transformations. Thanks to the efficient representation with integer sets and maps, POM can be easily extended to support more customized transformations.


\vspace{-0.1cm}
\subsection{MLIR affine dialect with HLS attributes }
The polyhedral IR is lowered to the MLIR affine dialect with HLS attributes, where hardware optimizations are performed. The affine dialect provides abstractions for affine operations and can be naturally mapped from the polyhedral AST. Besides, the affine dialect provides explicit loop structures required for hardware optimizations, which the abstract polyhedral IR lacks. This makes it a suitable IR to insert HLS pragma-related information as attributes for code generation. 

\noindent \textbf{Mapping from polyhedral AST to affine dialect}: Given the polyhedral IR, which consists of an annotated AST with hardware optimization information, POM maps different types of nodes in the AST to corresponding operations described in affine dialect, as shown in Fig. \ref{fig:polyhedralir}(d). For example, a \texttt{for-node} within the AST signifies a \textit{for} loop in the affine dialect (\ding{172}\ding{174}\ding{175}\ding{176}). It captures essential loop attributes such as the lower bound, upper bound, and iteration step size. Similarly, a \texttt{user-node} within the AST represents user-defined statements in the \texttt{compute} (\ding{177}\ding{178}), which is reserved in POM DSL. 
POM is equipped with a recursive method to parse the statements and data accesses described in POM DSL and transform them into correct affine dialect representations.
By implementing automated translation from polyhedral IR to MLIR affine dialect, POM bridges the gap between the powerful polyhedral model and MLIR. Therefore, it can freely interact with other excellent works in MLIR's ecosystem while fully exploiting the advantages of the polyhedral model. 

\noindent \textbf{Implementation of hardware optimization}:
Upon completing the mapping, hardware optimizations attached to the nodes are eventually performed. POM provides a set of FPGA-specific hardware optimizations that are conducted at the affine dialect by inserting HLS pragma-related attributes into the corresponding code hierarchy. For example, by specifying a scheduling primitive \texttt{S3.pipeline(i, 1)} in DSL, a related pipeline attribute is attached to the loop level i with an initiation interval of one in the affine dialect in Fig. \ref{fig:polyhedralir}(d), indicating the specific \textit{loop pipelining} pragma-type operation. After manipulations on the affine dialect, the fully optimized IR is sent to the back-end to generate synthesizable HLS C code, where all of the attributes are translated to HLS pragmas. 




\section{Design space exploration}\label{sec-auto-dse}
POM enables a wide range of transformation and optimization methods, all of which form an exponentially increasing design space. To reduce effort, POM provides a two-stage DSE engine to automatically search for design choices that produce high-quality FPGA accelerators.

\subsection{Dependence-aware code transformation}
Dependence-aware code transformation is performed at first to alleviate tight loop-carried dependencies and facilitate parallelism as much as possible.
As is introduced, the input function is represented as a data dependence graph where each node denotes a loop. The DSE engine traverses the graph and our dependence analysis tool checks the loop-carried dependence to give hints for loop transformations. 
More specifically, if any loop-carried dependence is captured inside the loop, loop interchange will be considered since it effectively changes the dependence distance. We further consider the situation where two conflicting loop interchange strategies are proposed by the tool. In these cases, we need to leverage other transformations such as loop splitting and loop skewing to restructure the code. To make sure that data dependencies after transformation have been alleviated as much as possible, the dependence analysis tool will iteratively recheck loop-carried dependencies after each transformation and additional transformations are applied if needed. This iterative process will be terminated if every node no longer has the tight dependence issue or the number of iterations has reached its pre-defined bounds. Then the transformed code is well-prepared for the second stage. 

\begin{figure}
    \centering
    \includegraphics[width=0.37\textwidth]{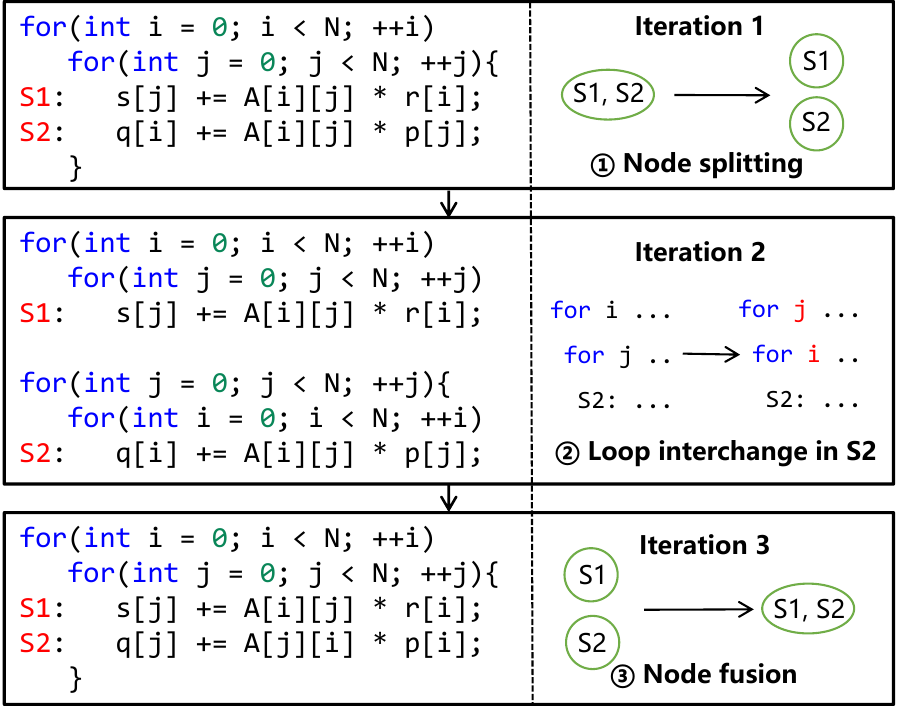}
    \caption{Illustration of dependence-aware code transformation. We use C code and graph nodes for illustration.} 
    \label{fig:bicg}
    \vspace{-0.4cm}
\end{figure}

We take Fig. \ref{fig:bicg} as an example. During the first iteration, the dependence analysis tool checks loop-carried dependence in both S1 and S2 and detects contradictory transformation strategies: S1 can execute efficiently without loop-carried dependence at the inner loop and hence tends to retain the current loop orders, while S2 has tight loop-carried dependence at the loop level j and tends to interchange i- and j- loop levels. 
To solve this issue, POM splits S1 and S2 into two independent loops, as shown in Fig. \ref{fig:bicg}\ding{172}. Then the dependence analysis tool continues iteratively rechecking dependencies inside each node, performing loop interchange to S2 in the second iteration (\ding{173}) and conservatively fuse S1 and S2 together in the third iteration (\ding{174}). After a series of transformations based on iterative dependence analysis, the potential parallelism of functions can be fully exploited.

\begin{table*}
\scriptsize
\renewcommand\arraystretch{1.1}
\caption{Evaluation and comparison on typical HLS benchmarks. The vector [$m$, $n$] denotes tiling sizes at different loop levels. 2MM and 3MM contain multiple loops with a sequence of tiling vectors.}
\label{tab:synthesis}
\begin{tabularx}{\linewidth}
{C{1.2cm}|C{1.1cm}|C{0.5cm}|C{0.85cm}|C{1.15cm}|C{1.40cm}|C{1.45cm}|C{0.6cm}|C{0.8cm}|C{2.75cm}|C{0.5cm}|C{0.75cm}}
\toprule[1pt]
\multirow{2}{5em}{\textbf{Benchmark}}& \multirow{2}{7em}{\textbf{Framework}} & \multirow{2}{*}{\begin{tabular}[c]{@{}c@{}}\textbf{Prob.}\\ \textbf{Size}\end{tabular}}& \multirow{2}{4em}{\textbf{Speedup}}  & \multirow{2}{*}{\begin{tabular}[c]{@{}c@{}}\textbf{DSP}\\ \textbf{(Util.\%)}\end{tabular}} & \multirow{2}{*}{\begin{tabular}[c]{@{}c@{}}\textbf{FF}\\ \textbf{(Util.\%)}\end{tabular}}& \multirow{2}{*}{\begin{tabular}[c]{@{}c@{}}\textbf{LUT}\\ \textbf{(Util.\%)}\end{tabular}} & \multirow{2}{*}{\begin{tabular}[c]{@{}c@{}}\textcolor{black}{\textbf{Power}}\\ \textcolor{black}{\textbf{(W)}}\end{tabular}} &  \multirow{2}{*}{\begin{tabular}[c]{@{}c@{}}\textbf{Achieved}\\ \textbf{II}\end{tabular}}& \multirow{2}{*}{\begin{tabular}[c]{@{}c@{}}\textbf{Achieved tile sizes and}\\ \textbf{ unroll factors}\end{tabular}}& \multirow{2}{*}{\begin{tabular}[c]{@{}c@{}}\textbf{Paral-}\\ \textbf{lelism}\end{tabular}} & \textcolor{black}{\textbf{DSE Time(s)}}\\
\midrule[1pt]

\multirow{3}{3.8em}{\textbf{GEMM}} 
 & \textbf{POLSCA} &  4096 & $2.3\times$ & 7 ($3\%$) & 4980 ($4\%$) & 7817 ($14\%$) & -&248  & - & - & - \\
& \textbf{ScaleHLS}& 4096  & $576.1\times$ & 214 ($97\%$) & 41616 ($39\%$) & 42676 ($80\%$) & \textcolor{black}{0.767} & 4  & [2, 4, 16]  & 32 & \textcolor{black}{24.4} \\
& \textbf{POM} & 4096 & $575.9\times$  & 166 ($75\%$) & 23067 ($21\%$) & 30966 ($58\%$) & \textcolor{black}{0.459} & 1  & [1, 2, 16] & 32 &  \textcolor{black}{11.4} \\
\hline
\multirow{3}{3em}{\textbf{BICG}} 
&\textbf{POLSCA} & 4096 &  $2.1\times$  & 5 ($2\%$) & 4665 ($4\%$)& 8150 ($15\%$) & - & 161 & - & - & - \\
& \textbf{ScaleHLS} & 4096  & $41.7\times$   & 32 ($14\%$) & 17326 ($16\%$)& 10386 ($19\%$) & \textcolor{black}{0.176} & 43 & [16, 8] & 3.0 & \textcolor{black}{6.3} \\
&\textbf{POM} & 4096 & $224.0\times$ & 160 ($72\%$) & 27189 ($25\%$) &43823 ($82\%$) & \textcolor{black}{0.782} & 2 & [1, 32] & 16 & \textcolor{black}{5.7} \\
\hline
\multirow{3}{6em}{\textbf{GESUMMV}} 
&\textbf{POLSCA} & 4096 &  $1.4\times$ & 8 ($3\%$) &  6112 ($5\%$) &  10979 ($20\%$) 
& - &161  & - & - & -  \\
&\textbf{ScaleHLS} & 4096  & $199.1\times$ & 158 ($72\%$)  & 49838 ($46\%$) & 35848 ($67\%$) & \textcolor{black}{0.643} & 9 & [8, 16] & 14.2 & \textcolor{black}{6.4} \\
&\textbf{POM}& 4096 & $223.2\times$  & 160 ($72\%$) & 19409 ($18\%$) & 27595 ($51\%$) & \textcolor{black}{0.490} & 1 & [1, 16] & 16 &  \textcolor{black}{4.7}  \\
\hline
\multirow{3}{3em}{\textbf{2MM}} 
&\textbf{POLSCA} & 4096 &  $2.0\times$  & 8 ($3\%$)  &7137 ($6\%$) &11355 ($21\%$) & -& 248 & - & - & - \\
&\textbf{ScaleHLS}& 4096  & $31.0\times$ & 166 ($75\%$)  & 34912 ($32\%$) &45419 ($85\%$) & \textcolor{black}{0.462} & 4, 1 & [1, 8, 16], [1, 1, 1] & 1.9 & \textcolor{black}{77.2}   \\
&\textbf{POM}& 4096 & $510.1\times$  & 166 ($75\%$)  & 28039 ($26\%$)& 38577 ($72\%$)&  \textcolor{black}{ 0.537}& 1 & [1, 2, 16], [1, 2, 16] & 32 &  \textcolor{black}{24.5}  \\
\hline
\multirow{3}{3em}{\textbf{3MM}} 
&\textbf{POLSCA} & 4096 &  $1.8\times$ & 10 ($4\%$) & 10128 ($9\%$) & 16831 ($31\%$) & -& 256 & - & - & - \\
&\textbf{ScaleHLS}& 4096  & $40.1\times$ & 115 ($52\%$) & 34522 ($32\%$)& 38007 ($71\%$)& \textcolor{black}{ 0.599} & 1, 3, 3 & [1, 1, 1], [1, 8, 8], [1, 8, 8] & 2.7 & \textcolor{black}{56.8}  \\
&\textbf{POM}& 4096 & $335.4\times$ & 160 ($72\%$) & 21928 ($20\%$) &32995 ($62\%$) & \textcolor{black}{ 0.513} & 1 & [1, 2, 8], [1, 2, 8], [1, 2, 8] & 16 & \textcolor{black}{31.3} \\
\bottomrule[1pt]
\end{tabularx}
\end{table*}

\subsection{Bottleneck-oriented code optimization}
Since loop-carried dependencies are alleviated as much as possible during the first stage, we can focus on exploring parallelism by evaluating the combination of different loop-tiling strategies and HLS hardware optimizations. 
The core idea is to prioritize performance optimization (i.e., latency reduction) of the bottleneck node in the critical path.

At first, POM estimates the latency of each node in the dependence graph and gets the latency of each path, using the in-house model from \cite{ye2022scalehls}\cite{zhao2017comba}, which has been integrated into MLIR. After estimation, paths are ordered by their latency, and the critical path with the longest latency is selected to be optimized first. POM DSE further chooses the node with the longest latency in the critical path and performs a series of optimization strategies on it, varying types and factors of loop tiling and HLS optimizations. The set of types and factors are determined before the search and users can specify suitable groups of strategies and parameters. In this paper, we specify a range of strategies for loop tiling and HLS optimizations to cover different parallelism degrees. When optimizing the bottleneck node in the critical path, the DSE engine increases the parallelism degree gradually and sets the strategies correspondingly. Once the current node or path is not the bottleneck, the algorithm will switch to the new bottleneck for optimization and repeat the process. An exit mechanism is considered to avoid continuously optimizing the same node if it is always the longest one: the optimization will stop if the current node achieves its maximum parallelism degree or consumes resources that exceed resource constraints. We use a list to store all the nodes to be optimized. If the exit mechanism is triggered for one node, this node will be removed from the list. The DSE terminates when the optimization list is empty.

\section{Evaluation}

\subsection{Experimental setup}

Xilinx Vitis HLS and Vivado 2022.1 are utilized for HLS synthesis and hardware implementation. The reported performance and resource statistics are collected from HLS synthesis reports and 
\textcolor{black}{power statistics are obtained from implementation reports.} The target device is Xilinx XC7Z020 FPGA, containing 220 DSPs, 53,200 LUTs, 106400 FFs, and 4.9 Mb memories. 
\color{black}All the benchmarks are tested at the 100MHz target frequency with data types of 32-bit floating-point.
\color{black}

\subsection{Evaluation on typical HLS benchmarks}
\label{sec: VIIA}

We compare POM with state-of-the-art HLS frameworks on MLIR, POLSCA \cite{2022fplpolsca} and ScaleHLS \cite{ye2022scalehls}. Evaluation is performed on the same typical HLS benchmarks \cite{polybenchcode} with large problem sizes (4096), namely GEMM, BICG, GESUMMV, 2MM, and 3MM. Table \ref{tab:synthesis} compares latency speedups, resource usage, power, the achieved II and tile sizes, and the parallelism degree of the accelerators generated by the three frameworks, correspondingly. \color{black} We adopt the same target clock (10ns) as reported in ScaleHLS \cite{ye2022scalehls} for comparison. \color{black}
The DSE time costs are also compared in the last column.
\color{black}

\begin{figure}
    \centering
    \includegraphics[width=0.48\textwidth]{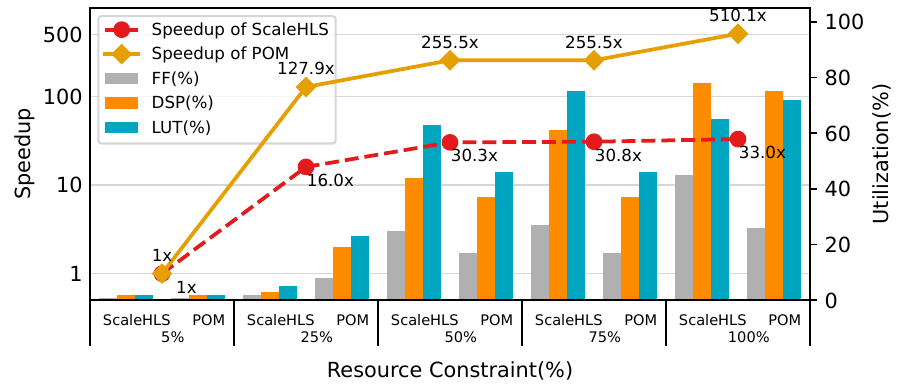}
    \caption{Speedup and resource utilization of \textit{2MM}.}
     \vspace{-0.5cm}
    \label{fig:resource}
\end{figure}


\begin{figure*}
    \centering
    \includegraphics[width=0.99\textwidth]{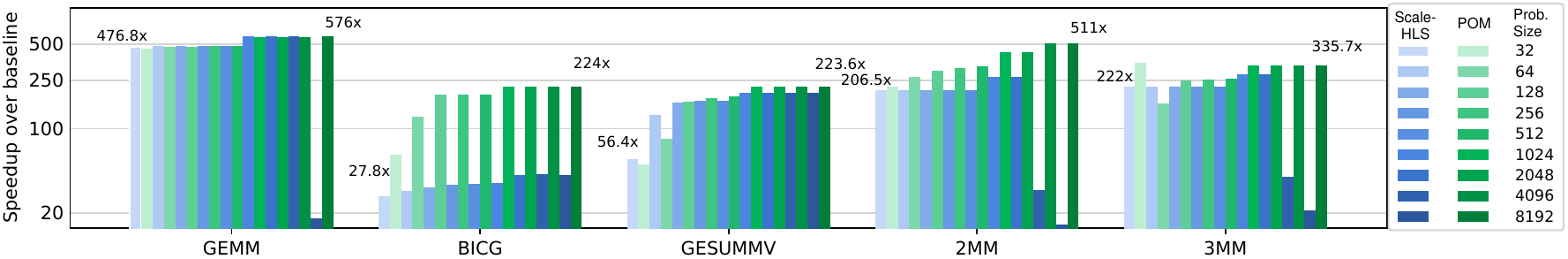}
    \caption{\textcolor{black}{Comparison and evaluation with different problem sizes on typical HLS benchmarks.  }}
    \label{fig:prob_size}
    \vspace{-0.4cm}
\end{figure*}

The \textit{latency speedup} is computed by dividing the latency (\textit{\#clock cycles}) of the original C code without any optimization by the latency (\textit{\#clock cycles}) of the optimized HLS C code. Experimental results show that POM significantly improves the overall performance of the baseline, resulting in speedups ranging from $223.2\times$ to $575.9\times$ across all the benchmarks. The achieved \textit{II} reflects the degree of parallelism between successive iterations in a pipeline loop, with smaller values indicating greater parallelism. The achieved tile sizes and unroll factors denote the number of parallel copies of computation units being executed, with larger values indicating greater parallelism. Therefore, to quantify the attained parallelism, we compute the \textit{parallelism} degree by dividing the product of tile sizes by the achieved \textit{II}, with higher values indicating greater parallelism achieved by the optimized accelerators.

We can see that POLSCA achieves limited speedups. This is primarily due to the presence of loop-carried dependencies in the code generated by Pluto in POLSCA, which negatively impacts the level of \textit{parallelism} achieved. Additionally, when handling large problem sizes, POLSCA does not properly partition arrays, resulting in a further decline in performance. Since Pluto works as a black box, its tile sizes and unroll factors are unknown to users.

Compared to ScaleHLS, POM achieves better speedups on most benchmarks with a $6.46\times$ performance improvement on average.
This improvement mainly comes from three aspects. First, with our accurate dependence analysis at the dependence graph IR, tight loop-carried dependencies are detected and alleviated, resulting in better achieved \textit{II} and larger tile sizes/unroll factors. This greatly improves our parallelism degree. For example, POM improves the performance of BICG by 224x speedup with \textit{parallelism=16}, while ScaleHLS achieves 41.7x speedup with \textit{parallelism=3} due to tight dependence inside the loop. Second, by introducing the powerful polyhedral model into MLIR at the polyhedral IR level, more effective loop transformations can be conducted efficiently while ensuring the correctness of the code. This actually enlarges the design space and increases the possibility of finding a design choice with higher performance. Third, although the design space is enlarged, our two-stage DSE engine still explores the large design space efficiently and finds high-performance design choices successfully. \textcolor{black}{For example, 3MM consists of multiple loops in multiple paths. The DSE engine of POM prioritizes the optimization of the bottleneck loop and switches to other loops once a new bottleneck appears, leading to concurrent optimization for all the loops. In contrast, ScaleHLS optimizes some loops heavily without leaving additional optimization space for other loops. For example, the first loop of 3MM is not tiled and unrolled with factors $[1,1,1]$.}

For the DSE time cost, it takes a shorter DSE time for POM on all benchmarks. Although POM achieves a smaller but closer speedup compared to ScaleHLS for GEMM (0.99x), it takes 50.1\% time cost for POM to find this design. \color{black}{Note that the code generation process from MLIR to HLS C typically completes within 0.1s. Therefore, the DSE time can be considered as the toolchain's runtime.}
\color{black}The rest columns in Table \ref{tab:synthesis} present resource usage with utilization ratios on the same FPGA. POM fully utilizes available resources given the same constraints which are set to the total resources on the board. We vary resource constraints to different percentages of the total resources, run frameworks, and evaluate \textit{speedup} and \textit{resource utilization} of generated accelerators. Figure \ref{fig:resource} shows the results of 2MM and POM achieves higher performance given different resource constraints. \textcolor{black}{Besides resource costs, we also compare power consumption in Table \ref{tab:synthesis}. POM achieves superior or competitive performance speedup while consuming less power for GEMM, GESUMMV, and 3MM. For BICG and 2MM, POM achieves remarkable speedups of 5.37x and 16.45x with an increase in power consumption by a factor of 4.44x and 1.16x, respectively, demonstrating a better performance-per-watt compared to ScaleHLS. 
}

\color{black}
\subsection{Comparison with manual optimization}
To better evaluate the quality of the automatically generated design, we compare it to a design that is manually optimized, leveraging our expertise in FPGA optimization. BICG is used for this case study. A series of HLS pragmas and code rewriting is performed to improve the parallelism during manual optimization. The results are shown in Table \ref{tab:comparison-manual}. We can see that the FPGA design generated by POM achieves a 1.39x speedup compared to the manually optimized HLS design and consumes fewer resources on the same FPGA device. 

\begin{table}
\scriptsize
\centering
\renewcommand\arraystretch{1.2}
\caption{\textcolor{black}{Comparison with manual optimization on BICG.}
}
\label{tab:comparison-manual}
\begin{tabular}{c@{  }|@{ }c@{ }|c|@{ }c@{ }|@{  }c@{  }|@{   }c@{   }}
\toprule
\textbf{Design} & \textbf{Cycles} & \textbf{Speedup} & \textbf{DSP(Util.\%)} & \textbf{FF(Util.\%)} & \textbf{LUT(Util.\%)} \\
\toprule
\textbf{Unoptimized} & 234889217 & 1$\times$ & 10(4\%) & 1101(1\%) & 1618(3\%) \\
\textbf{Manual opt.} & 1458178 & 161.1x & 208(94\%) & 23454(22\%) & 50899(95\%) \\
\textbf{DSE opt.} & 1048588 &  224.0x & 160(72\%) & 27189(25\%) & 43823(82\%) \\
\bottomrule
\end{tabular}
\vspace{-0.5cm}
\end{table}

\begin{table*}
\scriptsize
\renewcommand\arraystretch{1.2}
\caption{Comparison on image processing and DNN applications.}
\label{tab:eval-image}
\begin{tabularx}{\linewidth}{C{1cm}C{1.1cm}|C{0.4cm}|C{0.7cm}C{0.5cm}C{0.5cm} |C{0.8cm}C{0.7cm}C{0.5cm}|C{1.0cm}C{0.9cm}C{0.7cm} |C{1.2cm}C{0.9cm}C{0.7cm}}
\toprule[1pt]
\multicolumn{2}{c|}{\multirow{2}{6em}{\textbf{Applications}}} & \multirow{2}{2em}{\textbf{Prob. Size}} & \multicolumn{3}{c|}{\textbf{Speedup}} & \multicolumn{3}{c|}{\textbf{DSP (Utilization\%)}} & \multicolumn{3}{c|}{\textbf{FF (Utilization\%)}} & \multicolumn{3}{c}{\textbf{LUT (Utilization\%)}}\\
 \multicolumn{2}{c|}{} & & \textbf{ScaleHLS} &\textbf{POM} & \textbf{$P/S$}  & \textbf{ScaleHLS} & \textbf{POM} & \textbf{$P/S$}   & \textbf{ScaleHLS}  & \textbf{POM} & \textbf{$P/S$}  & \textbf{ScaleHLS}  & \textbf{POM} & \textbf{$P/S$} \\
\midrule[1pt]

\multirow{2}{3em}{\textbf{Image}} & \textbf{EdgeDetect} & 4096 & $19.1\times$ & $344.0\times$ & $18.0$ &23($10\%$) & 183($83\%$)& $8.0$ & 10130($9\%$) & 30686($28\%$)& $3.0$ & 11438($21\%$) & 47872($89\%$)& $4.2$ \\
 \multirow{2}{10em}{\textbf{Processing}} & \textbf{Gaussian} & 4096 & $111.4\times$ & $312.0\times$ & $2.8$ & 87($39\%$) & 177($80\%$) & $2.0$& 33104($31\%$) & 51203($48\%$) & $1.5$ & 28849($54\%$) & 52751($99\%$) & $1.8$\\
& \textbf{Blur} & 4096 & $59.3\times$ & $356.0\times$ & $6.0$ & 18($8\%$) & 48($21\%$) & $2.7$ & 7300($6\%$) & 11378($10\%$) & $1.6$ & 7006($13\%$) & 14549($27\%$) & $2.1$\\
\midrule[1pt]
\multirow{2}{2em}{\textbf{DNN}} & \textbf{VGG-16}  & 512 &  $33.6\times$     & $86.8\times$  & $2.6$   &137($62\%$) &40($18\%$)& $0.3$ & 38498($36\%$)& 40127($37\%$) & $1.0$ & 53819($101\%$\textcolor[rgb]{1,0,0}{\ding{56}}) &52837($99\%$) & $1.0$ \\ 
& \textbf{ResNet-18}  & 512 & $50.8\times$     & $46.4\times$    &  $0.9$ & 212($96\%$) &30($13\%$) & $0.1$ & 57882($54\%$)&38174($35\%$) & $0.6$ & 87662($164\%$\textcolor[rgb]{1,0,0}{\ding{56}}) &52484($98\%$) & $0.6$\\
\bottomrule[1pt]
\end{tabularx}
\end{table*}

\begin{figure*}
    \centering
    \includegraphics[width=\textwidth]{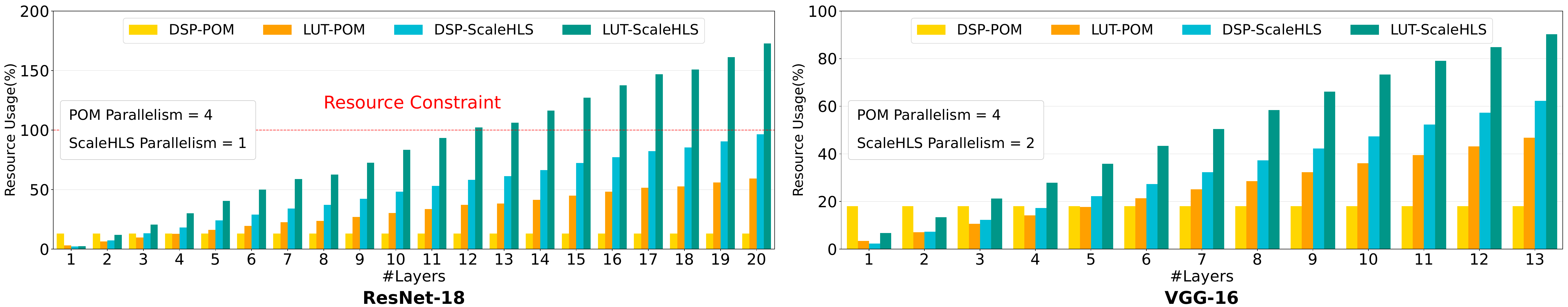}
    \caption{Accumulated resource usage for DNN workloads. We omit resource consumption of small loops such as initialization. }
    \label{fig:dnn}
\end{figure*}

\color{black}
\subsection{Evaluation on scalability with different problem sizes}
To evaluate the scalability of POM, we compare the performance of POM and ScaleHLS across various problem sizes on typical HLS benchmarks, as shown in Fig. \ref{fig:prob_size}. The problem sizes vary from 32 to 8192.
We can see that POM achieves superior performance for the majority of problem sizes. For problem sizes ranging from 32 to 2048, both POM and ScaleHLS exhibit stable performance improvement. However, as the problem size scales up to 4096 and 8192, there is a noticeable performance decline for ScaleHLS. For example, at a problem size of 8192, ScaleHLS encounters difficulties in generating efficient designs for benchmarks like GEMM, 2MM, and 3MM, providing only basic hardware optimizations such as loop pipelining. In contrast, POM 
continues to generate high-quality designs even as problem sizes expand to 8192. 
For certain benchmarks with very small problem sizes, such as GESUMMV with a problem size of 32, POM exhibits slightly inferior performance compared to ScaleHLS. This occurs because POM prioritizes the optimization of bottleneck loops while placing relatively less emphasis on simpler loops. These simple loops account for a higher proportion of latency when problem sizes (i.e., loop bounds) are small. The slightly lower performance is acceptable since the absolute latency difference is quite minor for small problem sizes.

\color{black}

\begin{table}
\scriptsize
\centering
\renewcommand\arraystretch{1.2}
\caption{Comparison on optimization for critical loops.}
\label{tab:strategy}
\begin{tabularx}{0.99\linewidth}{c|YY|YY|YY }
\toprule
\multirow{2}{4em}{\textbf{Benchmark}} & \multicolumn{2}{c|}{\textbf{Tile Size}} & \multicolumn{2}{c|}{\textbf{Achieved II}} & \multicolumn{2}{c}{\textbf{Parallelism}} \\
& \textbf{ScaleHLS} & \textbf{POM} & \textbf{ScaleHLS} & \textbf{POM}& \textbf{ScaleHLS} & \textbf{POM}\\
\midrule
\textbf{EdgeDetection} & [1,2,3] & [2,2,3]  & 12, 6 & 1 & 0.67 &  12 \\
\textbf{Gaussian} & [1,3,3] & [1,3,3]  & 3 & 1 & 3 & 9 \\
\textbf{Blur}  & [2,1,3] & [2,2,3]  & 3 & 1  & 2 & 12  \\
\bottomrule
\end{tabularx}
\vspace{-0.3cm}
\end{table}

\begin{table}
\scriptsize
\centering
\renewcommand\arraystretch{1.2}
\caption{Evaluation on complicated code patterns.}
\label{tab:comparison-stencil}
\begin{tabularx}{0.95\linewidth}{c|c|c|c|c}
\toprule
\textbf{Benchmark} & \textbf{Speedup} & \textbf{DSP(Util.\%)} & \textbf{FF(Util.\%)} & \textbf{LUT(Util.\%)} \\
\midrule
\textbf{Jacobi-1d} & {$47.6\times$ } & $14(6\%)$ &2794(2\%)& 3746\%(7\%) \\

\textbf{Jacobi-2d} & {$136.0\times$ } & $44(20\%)$ &8542(8\%)& 13178\%(24\%) \\
\textbf{Heat-1d} & {$22.9\times$ } & $20(9\%)$ &2466(2\%)& 3832\%(7\%) \\
\textbf{Seidel} & {$53.8\times$ } & $73(33\%)$ &25327(23\%)& 20892\%(39\%) \\
\bottomrule
\end{tabularx}
\vspace{-0.4cm}
\end{table}

\begin{figure*}
    \centering
    \includegraphics[width=\textwidth]{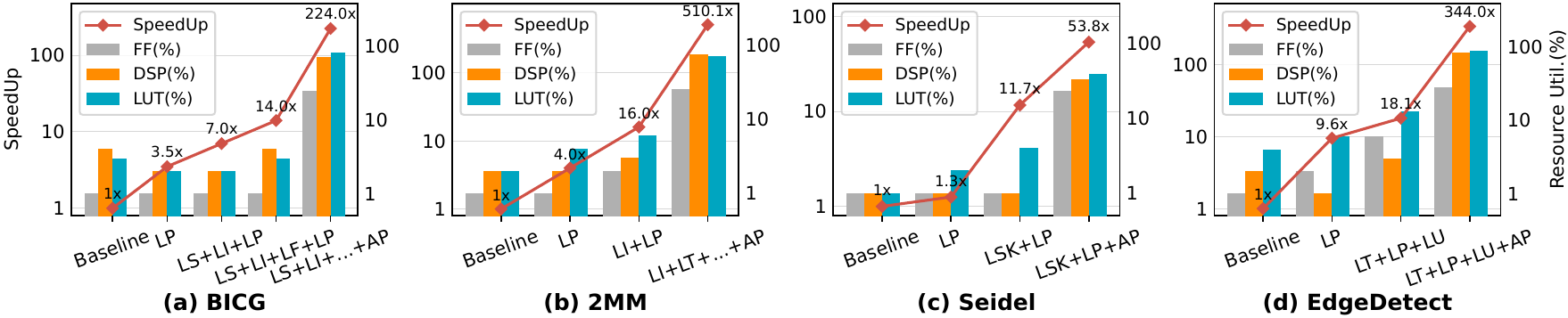}
    \caption{\textcolor{black}{Impact analysis of scheduling primitives. LI, LT, LS, LF, LSK, LP, LU and AP denote \textit{loop interchange}, \textit{loop tiling}, \textit{loop splitting}, \textit{loop fusion}, \textit{loop skewing}, \textit{loop pipelining}, \textit{loop unrolling}, and \textit{array partitioning}, respectively.}}
    \label{fig: impact analysis}
    \vspace{-0.4cm}
\end{figure*}

\subsection{Evaluation on image processing and DNN applications}

To demonstrate the ability of POM in accelerating complicated real-world applications, we evaluate the performance of ScaleHLS and POM on several image processing and DNN applications, including 
EdgeDetection \cite{baghdadi2019tiramisu}, Gaussian \cite{baghdadi2019tiramisu}, Blur \cite{ragan2017halide}, VGG-16 \cite{simonyan2014very} and ResNet-18 \cite{resnet15} on the same FPGA device. 
Table \ref{tab:eval-image} shows the detailed results, where \textbf{$P/S$} denotes the relative ratio between POM ($P$) and ScaleHLS ($S$). Compared to the baseline without optimization, POM significantly improves the performance with speedups from $46.4\times$ to $356.0\times$ \textcolor{black}{within minutes}. Compared to ScaleHLS, POM achieves $6.06\times$ speedup on average at similar time costs. 

For image processing, Table \ref{tab:strategy} compares the tile size, achieved \textit{II}, and \textit{parallelism} for critical loops. With accurate dependence analysis and an effective DSE engine, POM is able to achieve a higher parallelism degree compared to ScaleHLS. 

For DNN workloads, POM and ScaleHLS use different optimization strategies, as illustrated in Fig. \ref{fig:dnn}. \color{black}{Due to space limit, statistics for critical loops are visualized in the figure. Here critical loops refer to the nested loops with a loop level exceeding four in the neural network. 
For instance, ResNet-18 has 20 critical loops, including 17 convolution loops and 3 residual loops, and VGG-16 has 13 critical loops, all of which are convolution loops.
} 
\color{black}POM improves the parallelism of each loop and resources are reused between different layers. Given a fixed number of total resources, this resource reuse increases the available resources for each layer in POM, resulting in higher parallelism for each layer. Consequently, DNN layers in POM are executed in sequence but the \textit{parallelism} of each layer (i.e., 4) is maximized due to resource reuse. 
In contrast, ScaleHLS makes layers executed in a pipelined dataflow, and resources are not shared among different layers. 
The overall latency is equal to that of the bottleneck layer. Since each layer occupies resources, the parallelism of each loop is degraded (i.e., 1), influencing the overall performance greatly, especially for DNNs with large \#layers. The pipelined dataflow will stall due to unmatched computation paces between successive loops. 
Therefore, POM achieves a $2.6\times$ speedup on VGG-16 compared to ScaleHLS. For ResNet-18, POM achieves a relatively lower speedup ($0.9\times$) while consuming much fewer resources ($0.1\times$ DSPs and $0.6\times$ LUTs), satisfying the resource constraints of our FPGA device. The optimized designs from ScaleHLS are not feasible since their resource usage exceeds the total resources on the target FPGA.



\subsection{Evaluation on complicated data access patterns}
We extensively test four applications with more complicated code patterns, including Jacobi-1d, Jacobi-2d, Heat-1d, and Seidel. For example, Seidel is a stencil computation with complex data access patterns with tight loop-carried dependence. 
\textcolor{black}{For these benchmarks, ScaleHLS and POLSCA fail to find an optimization strategy that improves the performance greatly.} 
\textcolor{black}{In contrast, POM generates high-quality designs within seconds.} 
Experimental results show that POM continuously performs well on these benchmarks and improves the overall performance of the baseline by $22.9\times$ to $136.0\times$ ($65.08\times$ on average). The baseline is the original implementation without optimization. The reason for the improvement is that POM supports more useful loop transformations, such as \textit{loop skewing}. We also notice that resource utilization ratios are relatively small for these benchmarks. This is because their loop-carried dependence degrades the parallelism of loops, even though we have relieved the dependence to some extent. 

\color{black}{
\subsection{Impact analysis of scheduling primitives}
To understand the impact of different scheduling primitives, we conducted an ablation study on representative benchmarks in Fig. \ref{fig: impact analysis}.  Performance speedup and resource usage are presented. 
We notice that different benchmarks may benefit from different primitives due to their specific loop structures and data access patterns. For example, EdgeDetect gains 9.6$\times$ speedup from \textit{loop pipelining}, while the improvement of Seidel applied with the same optimization is limited. This is due to the loop-carried dependence inside the Seidel loop and the overall performance is improved significantly after loop skewing is applied. 
2MM benefits a lot from combinations of loop transformations and hardware optimizations because the parallelism is fully explored. These results highlight the necessity of integrating both loop transformations and hardware optimizations to fully exploit parallelism. }
\color{black}
\begin{figure}
    \centering
    \includegraphics[width=0.47\textwidth]{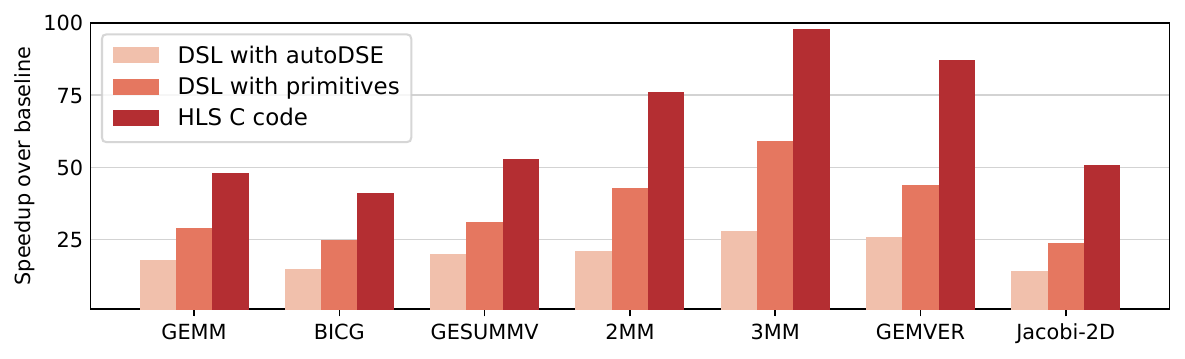}
    \caption{Comparison of lines of code (LoC).}
    \label{fig:loc}
    \vspace{-0.3cm}
\end{figure}

\begin{figure}
    \centering
    \includegraphics[width=0.47\textwidth]{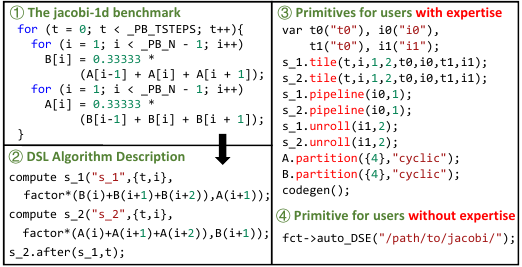}
    \caption{\textcolor{black}{Jacobi-1d described with POM DSL.}}
    \label{fig:dsl_case_study}
    \vspace{-0.5cm}
\end{figure}

\subsection{Evaluation of DSL expressiveness}
To evaluate the expressiveness of POM DSL, we first compare the number of lines of code (LoC) between the POM DSL and the equivalent HLS C code, as shown in Fig. \ref{fig:loc}.  \color{black}{Since the automatic DSE engine provided in POM frees programmers from explicitly setting scheduling primitives, we further consider two cases, namely DSL with the \texttt{autoDSE} primitive and DSL with \textit{manually-specified} primitives.}
\color{black}
Representative benchmarks with different complexities are chosen for evaluation. We describe these benchmarks with POM DSL, use the \texttt{autoDSE} primitive to automatically decide the scheduling, and utilize our tool to generate equivalent HLS C code. We also implement the same optimizations by manually setting the primitives in POM DSL. This ensures that the performance of these three kinds of code remains the same. Compared to HLS C code, POM DSL is able to express and optimize the same algorithm with much fewer lines of code. It takes less than one-third of the code for DSL with \texttt{autoDSE} to represent benchmarks with multiple loops such as 3mm. These results prove that POM DSL saves great engineering efforts with clear definitions of computations and arrays and efficient scheduling primitives.


\color{black}{To demonstrate the DSL usage, we also conduct a case study with the stencil benchmark, Jacobi-1d \cite{polybenchcode}, in Fig. \ref{fig:dsl_case_study}\ding{172}. POM utilizes \texttt{compute} and \texttt{after} to represent the nested loop (\ding{173}). For users with expertise, POM provides various scheduling primitives to help users explore different designs quickly (\ding{174}). For users lacking FPGA expertise, they can utilize the \texttt{autoDSE} primitive (\ding{175}) to automatically generate high-quality designs without explicitly specifying any other primitives. Note that the \texttt{autoDSE} primitive in \ding{175} is able to generate the same design as \ding{174}.
}
\color{black}

\section{Conclusion}
This paper proposes POM, an end-to-end optimizing framework on MLIR, to generate high-quality FPGA-based accelerators automatically. 
Experimental results show that accelerators generated by POM achieve significant speedup compared to SOTA. 
The whole compilation stack of POM will be \textbf{open-sourced} at \url{https://github.com/sjtu-zhao-lab/pom}.

\section{Acknowledgements}

This work is partially sponsored by the National Natural Science Foundation of China (62102249, 62232015, 61832006) and Shanghai Pujiang Program (21PJ1408200).


\clearpage

\end{document}